# Combinatorial tuning of electronic structure and thermoelectric properties in $Co_2MnAl_{1-x}Si_x$ Weyl semimetals

Rajkumar Modak[1,a)], Kazuki Goto[1], Shigenori Ueda[1], Yoshio Miura[1], Ken-ichi Uchida[1,2,3] and Yuya Sakuraba[1,4,a)]

AFFILIATIONS

[1]National Institute for Materials Science, Tsukuba 305-0047, Japan

[2]Institute for Materials Research, Tohoku University, Sendai 980-8577, Japan

[3]Center for Spintronics Research Network, Tohoku University, Sendai 980-8577, Japan

[4]PRESTO, Japan Science and Technology Agency, Saitama 332-0012, Japan.

[a)]Authors to whom correspondence should be addressed: MODAK.Rajkumar@nims.go.jp, SAKURABA.Yuya@nims.go.jp

ABSTRACT

A tuning of Fermi level ($E_F$) near Weyl points is one of the promising approaches to realize large anomalous Nernst effect (ANE). In this work, we introduce an efficient approach to tune $E_F$ for the $Co_2MnAl$ Weyl semimetal through a layer-by-layer combinatorial deposition of $Co_2MnAl_{1-x}Si_x$ (CMAS) thin film. A single-crystalline composition-spread film with $x$ varied from 0 to 1 was fabricated. The structural characterization reveals the formation of single-phase CMAS alloy throughout the composition range with a gradual improvement of $L2_1$ order with $x$ similar to the co-sputtered single layered film, which validates the present fabrication technique. Hard X-ray photoemission spectroscopy for the CMAS composition-spread film directly confirmed the rigid band-like $E_F$ shift of approximately 0.40 eV towards the composition gradient direction from $x = 0$ to 1. The anomalous Ettingshausen effect (AEE), the reciprocal of ANE, has been measured for whole $x$ range using a single strip along the composition gradient using the lock-in thermography technique. The similarity of the $x$ dependence of observed AEE and ANE signals clearly demonstrates that the AEE measurement on the composition spread film is an effective approach to investigate the composition dependence of ANE of Weyl semimetal thin films and realize the highest performance without fabricating several films, which will accelerate the research for ANE-based energy harvesting.

**INTRODUCTION**

A thermoelectric generator (TEG) using the Seebeck effect (SE) is one of the promising energy harvesting technologies which can generate electricity from everyday waste heat. During past many decades, researches on developing TEG using SE has been extensively conducted, but till now various remaining problems have to be solved.[1,2] One of the major drawbacks in SE is the direction of thermoelectric voltage which appears in parallel to the temperature gradient. Thus, SE-based module usually has a complicated structure in which the multiple thermopiles must be placed and connected at bottom and top alternatively in serial, causing high cost, low flexibility, and poor mechanical endurance. One possible solution to overcome these issues is developing a TEG based on the anomalous Nernst effect (ANE) which generates thermoelectric voltage in the perpendicular direction to a temperature gradient and a magnetization.[3] This transverse thermoelectric generation enables us to realize a much simpler laterally connected thermopile structure to increase the output thermoelectric voltage in a TEG.[4,5] These unique advantages have stimulated studies on ANE not only for gaining a fundamental understanding of the phenomenon but also for practical thermoelectric applications.[6-19]

At present, studies on ANE are still very less as compared to those on SE, and the reported thermopowers of ANE at around room temperature are usually very small to be implemented for practical device applications. Thus, an exploration of the materials showing high thermopower of ANE is clearly needed. Recent theoretical and experimental investigations suggested that the presence of intense Berry curvature (BC) in the vicinity of the Fermi energy ($E_F$) can potentially enhance the intrinsic anomalous Hall effect (AHE) and ANE.[20-25] Very recently, the large anomalous Nernst coefficient ($S_{ANE}$) of ~ 6 µV/K was reported in $Co_2MnGa$ at room temperature,[26,27] which is a member of the $Co_2YZ$-based full Heusler family having strong BC near $E_F$ originating from the Weyl points.[28,29] This $S_{ANE}$ value is almost one order of magnitude larger than that for conventional ferromagnetic materials, which opens a new possibility of enhancing ANE in these alloys. Sumida *et al.* has recently observed spin-polarized Weyl points near $E_F$ in $Co_2MnGa$ epitaxial films having different composition ratios and found that the magnitude of $S_{ANE}$ strongly depends on the position of BC with respect to $E_F$.[30] As an evidence of the importance of $E_F$ tuning, the magnitude of $S_{ANE}$ of 6.2 µV/K observed for $E_F$ tuned $Co_2MnGa$ film by Sumida *et al.* is nearly three times larger than that for the similar $Co_2MnGa$ epitaxial film reported earlier.[31] Therefore, to investigate the effect of theoretically predicted intrinsic BC and realize large $S_{ANE}$, tuning the position of $E_F$ is essential. Recently, the tuning of $E_F$ with atomic substitution has been performed in $Co_2MnAl_{1-x}Si_x$ (CMAS) alloys.[32-34] The replacement of Al by Si shifts the $E_F$ to higher energy and enhances the spin-polarization, AHE, and ANE for optimal CMAS composition. However, the tuning of $E_F$ by making individual specimens and performing systematic measurements for each are time-consuming and also cause a risk missing the best property of the material because of the unavoidable discontinuous change of the $E_F$. Therefore, more facile approach to achieve continuous $E_F$ tuning is strongly desired. Combinatorial deposition is one of the most effective tools which allows to fabricate composition spread films with single atom substitution from 0 to 100% on a single film. In this study, we fabricated a CMAS composition-spread thin film using the combinatorial deposition technique and experimentally demonstrated the systematic tuning of $E_F$ and other transport and thermoelectric properties with composition. Through the measurement of the anomalous Ettingshausen effect (AEE), which is the reciprocal effect of ANE, we employed a much faster and easier technique to effectively optimize the composition suitable for the transverse thermoelectric conversion.

**EXPERIMENTAL DETAILS**

A 001-oriented epitaxial CMAS composition-spread film with a thickness of 50.4 nm and a composition of $0 \leq x \leq 1$ was fabricated on a single-crystalline MgO (001) substrate using DC/RF magnetron sputtering with a base pressure of $< 6.0 \times 10^{-6}$ Pa and a process Ar gas pressure of 0.4 Pa. Before deposition, the MgO substrate was flashed at 790°C and then the film was deposited at 600°C substrate temperature. The CMAS composition-spread film with composition variation over a length of 7.0 mm was prepared from $Co_2MnAl$ (CMA) and $Co_2MnSi$ (CMS) alloy targets using the following deposition sequence: (1) deposition of a wedge-shaped CMA layer using a linear moving shutter with deposition rate 0.042 nm/sec and shutter speed 0.53 mm/sec, (2) rotation of the substrate by 180º, and (3) deposition of a wedge-shaped CMS layer using the linear moving shutter with deposition rate 0.026 nm/sec and shutter speed 0.33 mm/sec, where the total thickness of the CMAS film after completing (1) to (3) was designed to be 0.56 nm corresponding to the lattice constant of $L2_1$-ordered $Co_2MnSi$. The sequence was repeated 90 times to get around 50.4-nm-thick film [see Fig. 1(a)]. The film was capped with a 2-nm-thick Al film to prevent oxidation. The compositions in $x = 0$ and 1 regions in the CMAS composition-spread film were measured by X-ray fluorescence spectroscopy using the standard $Co_2MnAl$ and $Co_2MnSi$ films whose compositions were strictly premeasured by inductively coupled plasma mass spectrometry. The crystal structure and atomic ordering were investigated by X-ray diffraction with a Cu $K_\alpha$ X-ray source. The measurement was performed at different positions on the CMAS film along the composition gradient at an interval of 1.0 mm using a 0.5 mm incident slit. To investigate the variation in electronic structures with composition, hard X-ray photoemission spectroscopy (HAXPES) measurements were performed at BL15XU of SPring-8.[35] Excitation X-ray of 6 keV with the focal size of 25 μm (vertical) × 35 μm (horizontal) in FWHM at the sample position was irradiated at different places on the CMAS film to probe the different compositions. The composition gradient of the film was along the vertical direction. Thus, the vertical X-ray size is sufficiently small for composition dependent measurements. The incidence angle of X-ray was set to 88º deg, which expands the footprint of X-ray on the film in the horizontal direction. An additional film with 10.0 mm composition variation length was fabricated at the identical condition for the HAXPES measurements. Horizontal linearly polarized X-ray was used to excite photoelectron and excited photoelectrons were detected by a hemispherical analyzer (VG Scienta R4000). The pass energy of the analyzer was set to 100 eV, and the total energy resolution was approximately

150 meV. All measurements were performed at room temperature in nearly normal emission geometry.[36]

We performed the first-principles calculation to analyze the observed valance band spectra by HAXPES. The first-principles electronic structure calculations were performed with the Vienna ab initio simulation package.[37,38] The spin-polarized generalized gradient approximation is adopted for the exchange and correlation terms.[39] The atomic core potential is described by the pseudopotential with the projector augmented wave method.[40,41] In the calculation of photoemission spectra for CMS and CMA, we take into account the effects of the photoionization cross sections for constituent elements and the electron life time by Lorentzian smearing.

To investigate the dependence of AHE and ANE on the composition ratio of Al to Si efficiently, the composition-spread CMAS film was patterned into the parallel aligned Hall bars as shown in Fig. 1(b) by photolithography and Ar ion etching processes. The width of each bars is designed to be 200 μm, thus the composition variation of Al:Si in one Hall bar is estimated to be about 2.8%. AHE was measured by applying a charge current of 1 mA in the $y$ direction and the external magnetic field in the $z$ direction and measured the anomalous Hall voltage along the $x$ direction [see Fig. 1(b)]. ANE was measured by applying a temperature gradient $\nabla T$ of 0.3 K/mm in the $x$ direction and the external field in the $z$ direction and the measured anomalous Nernst voltage in the $y$ direction [see Fig. 1(c)]. It should be mentioned that $\nabla T$ must be equal in all Hall bars with different CMAS compositions because $\nabla T$ in the CMAS film is determined by the thermal conductance of the thick (0.5 mm) MgO substrate. The $\nabla T$ distribution was precisely evaluated by the infrared camera with the black body coating on the sample to correct an emissivity, which is a technique established in earlier studies.[18,19,30,34] SE was measured by applying $\nabla T$ in the y direction and measured voltage in the same direction [see Fig.1(d)]. AEE measurement was performed for the CMAS film using the lock-in thermography (LIT) technique.[42-46] To measure AEE, the film was pattered in a strip shape of the 8.0 mm length and 0.4 mm width, where the length direction is along the composition gradient [see Fig. 1(e)]. During the AEE measurement, a square-wave-modulated AC charge current with the square-wave amplitude of 10 mA, frequency $f$ = 25 Hz, and zero DC offset was applied along the strip and an external magnetic field with the magnitude $\mu_0 H = \pm 0.1$ T was applied along the $y$ direction (see section I in supplementary material). The pure AEE contribution was extracted using the previously established procedures from the raw LIT images.[47-53] Note that all patterns shown in Fig.1 were made from the identical CMAS composition spread film.

**RESULTS AND DISCUSSION**

Figure 2(a) shows the out-of-plane X-ray diffraction (XRD) pattern for the CMAS epitaxial film grown on the MgO (001) substrate at 600 °C. The only presence of 002 and 004 peaks for $\theta$-$2\theta$ scan confirms the 001-oriented growth of the CMAS alloy on MgO (001) for all composition range. In order to measure the 111 superlattice peak, the sample stage was tilted to 54.7° from the normal direction. As shown in the inset of Fig. 2(a), the 111 peak is absent in the CMA side and gradually appears with increasing intensity while moving to the CMS side. The 002 and 111 superlattice peaks give information about the $B2$ and $L2_1$ atomic ordering in these alloys, respectively. The $B2$ is the ordering between Co and (Mn, Al/Si) site [see Fig. 2(b)] and the $L2_1$ is the structure having the ordering between Mn and Al/Si site [see Fig. 2(c)]. The presence of the 002 peak in the XRD patterns indicate $B2$ ordering structure for all the composition range. Similarly, the absence of the 111 peak at the CMA side signifies the absence of $L2_1$ ordering which is in accordance with previous results for CMA.[34] With the introduction of Si at the Al site, the 111 peak appears and gradually increases when moved towards the CMS side which signifies the formation of $L2_1$ ordering. The degree of order for $B2$ and $L2_1$ structures, defined as $S_{B2}$ and $S_{L2_1}$, have been evaluated using the following equations:[54]

$$S_{B2} = \sqrt{\frac{\left[I_{002}/I_{004}\right]_{\text{exp.}}}{\left[I_{002}/I_{004}\right]_{\text{cal.}}}}$$

$$S_{L2_1} = \frac{2}{3 - S_{B2}} \sqrt{\frac{\left[I_{111}/I_{004}\right]_{\text{exp.}}}{\left[I_{111}/I_{004}\right]_{\text{cal.}}}}$$

where $I_{002}$, $I_{004}$, and $I_{111}$ represent the integrated intensities of the 002, 004, and 111 peaks, respectively. The subscript exp. and cal. represent the values obtained from the experimental XRD pattern and simulated XRD pattern for ideal $L2_1$ structure, respectively. For the present analysis, we performed the XRD pattern simulation for $L2_1$-ordered CMAS alloys using the visualization for electronic and structural analysis (VESTA) (The parameters of the simulation can be seen in section II in the supplementary material), where we used the actual compositions measured for pure CMA and CMS regions and estimated a linear variation of composition from the CMA side to the CMS side, as shown in Fig. 2(d). The obtained $B2$ and $L2_1$ ordering parameters are summarized in Fig. 2(e). It is observed that the degree of $S_{B2}$ is nearly 1 in the whole range of $x$, indicating nearly perfect $B2$ ordering exists throughout the CMAS composition-spread film. In contrast, $S_{L2_1}$ is zero for CMA, and rapidly increases from zero to 0.8 with $x$ when $x$ is changed from 0.1 to 0.4. The alloy with $x > 0.4$ has strong $L2_1$ which slowly improves with Si enrich and a maximum $L2_1$ order of 0.9 was observed for CMS. The lattice constant $a$, as evaluated from the 004 peak position, is plotted as a function of $x$ in Fig. 2(f). The $a$ values for the CMA and CMS ends are found to be 5.726 and 5.646 Å, respectively, which are comparable with the reported values in literature.[34,54,55] The $a$ almost linearly decreases with increasing the Si content following the Vegard's law, indicating that Al atoms were substituted into the Si sites without phase separation and that the single-phase CMAS alloy film was formed in the whole composition range.

Figure 3(a) represents the valence band HAXPES spectra for the CMAS film at different compositional positions defined as P1 to P6 [see Fig. 3(a) inset]. The observed spectra for the CMS site are in well agreement with the previously reported HAXPES data in bulk and thin films,[56-59] as well as the theoretical prediction based on the density of states (DOS) for $L2_1$-ordered CMS alloy [see Figs. 3(c) and 3(d)]. The observed highest intensity at the binding energy ($E_B$) of ~1.26 eV just below $E_F$ for CMS is due to flat $d$ bands that belong to minority $t_{2g}$ states localized in the Co planes as well as localized Mn majority $e_g$ states, as already described in detail for bulk samples.[56-59] The shifting of the peak position towards the higher $E_B$ side signified the shifting of $E_F$ while moving to the CMA side. The total shifting of 0.40 eV was observed from CMA to CMS [see Fig. 3(b)]. This value is consistent with the expectation based on the DOS calculation as shown in Fig. 3(c) and (d), where the Fermi level is predicted to be shifted around 0.45 eV with the composition variation from CMA to CMS. This peak shift can be qualitatively explained by the rigid band picture of Co-based Heusler alloys. The total shifting of 0.45 eV corresponds to the difference of the number of the valence electron between CMA and CMS. The experimental observation clearly demonstrated a systematic shift of $E_F$ with changing Al/Si ratio for the single CMAS composition-spread film which clearly established the usefulness of the present combinatorial investigation approach.

The $x$ dependence of the anomalous Hall resistivity ($\rho_{yx}$) and the longitudinal resistivity ($\rho_{xx}$) has been measured for the CMAS composition-spread film with the 7.0 mm composition variation length by patterning the film into Hall bars as described in the experimental section. Figure 4(a) shows the out-of-plane magnetic field dependence of $\rho_{yx}$ for the CMAS film for various values of $x$. $\rho_{yx}$ increases with the field and saturates when the saturation magnetization is reached, as expected for a ferromagnetic material. Figures 4(b) and 4(c) show the measured $\rho_{yx}$ and $\rho_{xx}$ values as a function of $x$, respectively. The $\rho_{xx}$ has the largest (smallest) value of 480 μΩ-cm (40 μΩ-cm) for CMA (CMS). Similarly, the CMA film shows the largest $\rho_{yx}$ value of 22.3 μΩ-cm which monotonically decreases with increasing $x$ and reaches to 0.06 μΩ-cm for CMS. These tendencies are consistent with the pervious report on CMAS films.[34] Figure 4(d) represents the estimated anomalous Hall angle ($\theta_{AHE} = \rho_{yx}/\rho_{xx}$) values for the CMAS film as a function of $x$. The observed $\theta_{AHE}$ gradually decreases with increasing the Si percentage; CMA showed the largest magnitude of $\theta_{AHE}$ of 4.8%, while CMS showed nearly zero $\theta_{AHE}$. The observed difference in the composition dependence of $\rho_{xx}$, $\rho_{yx}$, and $\theta_{AHE}$ between the present CMAS film and the previous co-sputtered CMAS films in ref. 34 can be roughly explained by a difference in composition and total valence electron number $N_v$ of the CMA in these two films. Namely, $N_v$ is 27.9 in the $x = 0$ region of the present CMAS film and 27.5 in the CMA film in the previous study because the compositions of the former and latter are $Co_{52.1}Mn_{21.1}Al_{26.8}$ and $Co_{48.4}Mn_{24.5}Al_{27.1}$, respectively. Recent investigation on $Co_2MnGa$ Weyl semimetal thin films having a various off-stoichiometric composition clearly demonstrates that the transport properties strongly depend on $N_v$ calculated from their compositions.[30] The similar effect can also be expected in CMAS films. Since the CMA of the current CMAS film has larger $N_V$ by 0.4 than the previous CMA film, we can see a closer matching between the data in two CMAS films by shifting of $\rho_{xx}$, $\rho_{yx}$ and $\theta_{AHE}$ vs $x$ for the present film toward the $+x$ direction by ~0.4 (see Fig. S2 in supplementary material) although there is still a certain amount of the disagreement which originates from the factors that do not scale with only $N_V$ but each composition ratio of Co, Mn and Al/Si. For example, the obtainable degree of $L2_1$ order depends on not $N_V$ but the Al/Si composition ratio, which might be a reason for much larger $\rho_{xx}$ for the CMA with no $L2_1$-order in the present CMAS film than the $Co_2MnAl_{0.6}Si_{0.4}$ with a partial $L2_1$-order in the previous study regardless of the same $N_V$ (~ 27.9) in these films.

To investigate the thermoelectric property of the CMAS film, SE and ANE were measured. The $x$ dependence of the Seebeck coefficient ($S_{SE}$) for the film is shown in Fig. 5(b). The $S_{SE}$ is found to be -14 μV/K for CMA and reach a maximum value of -22 μV/K at around $x = 0.3$. With further increasing of $x$, $S_{SE}$ decreases and remains almost invariant for $x > 0.65$. Figure 5(a) represents the measured ANE voltage ($V_{ANE}$) as a function of the applied magnetic field measured at different positions of the CMAS film. The ANE voltage saturates around 1 T, which signifies that the signal is dominated by the ANE. The anomalous Nernst coefficient ($S_{ANE}$) is evaluated using the relation $\mathbf{E}_{ANE} = -S_{ANE}\mathbf{m} \times \nabla T$, where $\mathbf{m}$ is the normal vector along the magnetization. The $x$ dependence of $S_{ANE}$ is summarized in Fig. 5(c). The $S_{ANE}$ value is +1.3 μV/K for CMA and increased to a maximum value of +2.7 μV/K for $x \approx 0.08$. $S_{ANE}$ then gradually decreases with increasing $x$. Interestingly, the observed trends of $S_{SE}$ and $S_{ANE}$ with respect to $x$ are similar to the previous study based on the individual CMAS films but shifts to less $x$ (Si-poor) direction.[34] This shift can be explained by a difference in $N_V$ in a similar manner to the above-mentioned $\rho_{xx}$, $\rho_{yx}$ and $\theta_{AHE}$. One can see the closer matching of $S_{SE}$ vs $x$ and $S_{ANE}$ vs $x$ in the two films by shifting the data for the present film towards the $+x$ direction by ~0.4 (see Fig. S2 in supplementary material). However, the maximum $S_{ANE}$ of +2.7 μV/K is smaller than that in the previous report (+3.9 μV/K).[34] Here we would like to explain a reason for small $S_{ANE}$ in the present CMAS film. One can see from the band dispersion on the high symmetry line (Fig. S3 of the supplementary material) that the energetical position of the Weyl cones giving large Berry curvature is very close to the $E_F$ in $L2_1$-Co$_2$MnAl (0-0.2 eV below the $E_F$) whereas far below the $E_F$ (0.5-0.7 eV) in Co$_2$MnSi. That is an origin of the giant AHE observed in $L2_1$-ordered single-crystalline Co$_2$MnAl.[60] However, it is usually difficult to obtain $L2_1$-atomic order in Co$_2$MnAl, especially in a thin film form, because of its much less energetical stability of $L2_1$-structure compared to Co$_2$MnSi as we found in this study. The previous study reported that $B2$-disordered structure smears the band dispersion forming Weyl cones and reduces AHE and ANE.[34] Therefore, one promising approach for getting the effect of large Berry curvature on AHE and ANE in CMAS is to make a Co$_2$MnAl having an off-stoichiometric composition that shifts the $E_F$ toward lower energy level by hole doping ($N_V < 28.0$) and replaces Al with Si to not only shift the $E_F$ toward higher level and but also to improve $L2_1$-ordering. The CMA in the present CMAS film has the composition ratio giving less hole doping ($N_V = 27.9$) than that in the previous report ($N_V = 27.5$), which results in small $S_{ANE}$ because the optimum Si composition for maximum ANE is close to the CMA side where the degree of $L2_1$-ordering is small.

We have successfully demonstrated the advantage of preparing a single composition-spread film fabricated using the combinatorial sputtering technique over preparing several uniform films of different compositions to investigate the composition dependence of ANE. However, the direct measurement of ANE and other transport properties for such a composition-spread film requires complicated design which makes high-throughput and systematic investigations difficult. For example, there is a limitation of the number of Hall bars made along the composition gradient as shown in Figs. 1(c) and 1(d), which limits the minimum composition difference between two successive bars and also one-by-one thermopower measurements take long time. This limitation can be solved by the combination of the combinatorial sputtering method and LIT technique. The LIT imaging detection enables high-throughput material screening for ANE through the measurement of it's Onsager reciprocal: AEE. The AEE measurement using LIT requires only a simple strip along the composition gradient as shown in Fig. 1(e), thus enabling easy and efficient scanning for best composition region in a composition-spread film. Importantly, the AEE data is continuous along the composition gradient, which gives much accurate composition dependence of ANE/AEE. To demonstrate this, AEE for the CMAS composition spread film was measured along the composition gradient using the LIT technique.[42-46] Figures 6(a) and 6(b) show the lock-in amplitude ($A$) and phase ($\phi$) images of the temperature modulation due to the Joule heating (defined as $A_{\text{Joule}}$ and $\phi_{\text{Joule}}$) and AEE (defined as $A_{\text{AEE}}$ and $\phi_{\text{AEE}}$), respectively. The $A_{\text{Joule}}$ value is proportional to the local resistivity in our configuration because the charge current density $j_c$ is uniform and the heat loss from the film to the substrate is independent of the position.[61] The $A_{\text{Joule}}$ image clearly demonstrates the higher resistance value at the CMA side, as shown in the $x$ dependence of the $A_{\text{Joule}}/j_c^2$ in Fig. 6(c) (note that $A_{\text{Joule}}$ is proportional to $j_c^2$). In fact, the tendency is consistent with the $\rho_{xx}$ variation as shown in Fig. 6(c) ($\rho_{xx}$ value is taken from Fig. 4(c)). The $A_{\text{AEE}}$ image, obtained by extracting the $H$-odd component of the detected LIT images,[47-53] clearly shows that AEE is intense at the CMA side and nearly vanishes at the CMS side. The $A_{\text{AEE}}/j_c$ as a function of $x$ is shown in Fig. 6(d) along with the measured $S_{\text{ANE}}$ value for the current film from Fig. 5(c). The result shows that $A_{\text{AEE}}/j_c$ first increases with $x$, and then gradually decreases in a similar manner to $S_{\text{ANE}}$. Large AEE was observed between $n = 0.06$ to $0.12$, which is consistent with the ANE result as can be seen in Fig. 6(d). The above result clearly shows that the imaging measurement of the temperature modulation due to AEE is an effective method to find the best composition for ANE in a composition spread film. Here one should note that it is possible to quantitatively estimate $S_{\text{ANE}}$ from the AEE-induced temperature modulation based on the Onsager reciprocal relation: $\Pi_{\text{AEE}} = S_{\text{ANE}}T$, where the anomalous Ettingshausen coefficient $\Pi_{\text{AEE}}$ is proportional to $A_{\text{AEE}}\cdot\kappa/j_c$ with $\kappa$ being the thermal conductivity of the ferromagnetic material.

The detailed procedure of extracting $\Pi_{AEE}$ from the observed AEE data can be seen in the previous reports.[48,51] The very close overlap between the AEE and ANE results in Fig. 6(d) suggests that the variation of $\kappa$ with the composition in our CMAS film is very small, although we did not perform the direct measurement of $\kappa$. This work demonstrates the usefulness of the combination of the combinatorial sputtering method and LIT technique as high-throughput material screening for finding materials showing large ANE and AEE. By performing the direct measurements of $S_{ANE}$ as well as $\kappa$ only for the optimum composition determined by the high-throughput screening, the materials exploration for ANE and AEE will be efficient. Importantly, the LIT-based method can be used not only for ANE but also for other thermoelectric and thermo-spin effects.[61,62]

**CONCLUSION**

We have successfully fabricated a composition-spread single-crystalline $Co_2MnAl_{1-x}Si_x$ Heusler alloy film on a MgO substrate using layer-by-layer wedge shape deposition. The $x$ is varied from 0 to 1 (0 to 100 % Si substitution) on a single film. The XRD characterization confirmed a formation of single phase throughout the composition range with linear variation of the lattice constant, showing the almost linear composition gradient over the film. The systematic tuning of Fermi energy level for CMA Weyl semimetal has been experimentally verified using HAXPES measurements and a shift of 0.40 eV was observed for varying CMA to CMS. The electrical resistivity, AHE, ANE, and SE properties for the composition-spread film have been investigated systematically. The composition dependence of AEE, the reciprocal transverse thermoelectric effect of ANE, for the composition-spread CMAS film has also been investigated by means of the LIT technique. The AEE result is consistent with the ANE result, which reveals a much easier and accurate approach to study the thermoelectric property of composition spread films. This research will accelerate the material investigation and optimization of the best composition with large ANE/AEE, which is essential to realize applications based on the transverse thermoelectric effects.

## SUPPLEMENTARY MATERIAL

See the supplementary material for the details on the procedure for extracting pure AEE signal, XRD simulation data, figure showing correlation of transport and thermoelectric properties of CMAS composition spread film with ref. 34 in terms of $N_V$ and band structures for $Co_2MnAl$ and $Co_2MnSi$ alloys.

## ACKNOWLEDGEMENTS
The authors thank R. Iguchi for valuable discussions and B. Masaoka and N. Kojima for technical supports. This work was supported by CREST "Creation of Innovative Core Technologies for Nano-enabled Thermal Management" (Grant No. JPMJCR17I1) and PRESTO "Scientific Innovation for Energy Harvesting Technology" (Grant No. JPMJPR17R5) from the Japan Science and Technology Agency and NEDO "Mitou challenge 2050" (Grant No. P14004). The HAXPES measurements were performed under the approval of the NIMS Synchrotron X-ray Station (Proposal Nos. 2020A4604 and 2020A4606).

## DATA AVAILABILITY
The data that support the findings of this study are available from the corresponding authors upon reasonable request.

## REFERENCES
[1] Q. Zhang, Y. Sun, W. Xu, and D. Zhu, Adv. Mater. **26**, 6829 (2014).
[2] K. Koumoto, R. Funahashi, E. Guilmeau, Y. Miyazaki, A. Weidenkaff, Y. Wang, and C. Wan, J. Am. Ceram. Soc. **96**, 1 (2013).
[3] T. Miyasato, N. Abe, T. Fujii, A. Asamitsu, S. Onoda, Y. Onose, N. Nagaosa, and Y. Tokura, Phys. Rev. Lett. **99**, 086602 (2007).
[4] Y. Sakuraba, K. Hasegawa, M. Mizuguchi, T. Kubota, S. Mizukami, T. Miyazaki, and K. Takanashi, Appl. Phys. Express **6**, 033003 (2013).
[5] Y. Sakuraba, Scr. Mater. **111**, 29 (2016).

[6]K. Hasegawa, M. Mizuguchi, Y. Sakuraba, T. Kamada, T. Kojima, T. Kubota, S. Mizukami, T. Miyazaki, and K. Takanashi, Appl. Phys. Lett. **106**, 252405 (2015).

[7]K. Uchida, T. Kikkawa, T. Seki, T. Oyake, J. Shiomi, Z. Qiu, K. Takanashi, and E. Saitoh, Phys. Rev. B **92**, 094414 (2015).

[8]Y. P. Mizuta and F. Ishii, Sci. Rep. **6**, 28076 (2016).

[9]D. J. Kim, K. D. Lee, S. Surabhi, S. G. Yoon, J. R. Jeong, and B. G. Park, Adv. Funct. Mater. **26**, 5507 (2016).

[10]M. Mizuguchi, and S. Nakatsuji, Sci. Technol. Adv. Mater. 20, 262 (2019).

[11]H. Narita, M. Ikhlas, M. Kimata, A. A. Nugroho, S. Nakatsuji, and Y. Otani, Appl. Phys. Lett. **111**, 202404 (2017).

[12]S. Tu, J. Hu, G. Yu, H. Yu, C. Liu, F. Heimbach, X. Wang, J. Zhang, Y. Zhang, A. Hamzić, K. L. Wang, W. Zhao, and J.-P. Ansermet, Cit. Appl. Phys. Lett **111**, 222401 (2017).

[13]T. C. Chuang, P. L. Su, P. H. Wu, and S. Y. Huang, Phys. Rev. B **96**, 174406 (2017).

[14]Z. Yang, E. A. Codecido, J. Marquez, Y. Zheng, J. P. Heremans, and R. C. Myers, AIP Adv. **7**, 095017 (2017).

[15]R. Ando and T. Komine, AIP Adv. **8**, 056326 (2018).

[16]P. Gautam, P. R. Sharma, Y. K. Kim, T. W. Kim, and H. Noh, J. Magn. Magn. Mater. **446**, 264 (2018).

[17]S. Isogami, K. Takanashi, and M. Mizuguchi, Appl. Phys. Express **10**, 073005 (2017).

[18]H. Nakayama, K. Masuda, J. Wang, A. Miura, K. Uchida, M. Murata, and Y. Sakuraba, Phys. Rev. Mater. **3**, 114412 (2019).

[19]W. Zhou and Y. Sakuraba, Appl. Phys. Express **13**, 043001 (2020).

[20] D. Xiao, M.-C. Chang, and Q. Niu, Rev. Mod. Phys. **82**, 1959 (2010).

[21] S. Y. Huang, W. G. Wang, S. F. Lee, J. Kwo, and C. L. Chien, Phys. Rev. Lett. **107**, 216604 (2011).

[22] M. Ikhlas, T. Tomita, T. Koretsune, M.-T. Suzuki, D. Nishio-Hamane, R. Arita, Y. Otani, and S. Nakatsuji, Nat. Phys. **13**, 1085 (2017).

[23]X. Li, L. Xu, L. Ding, J. Wang, M. Shen, X. Lu, Z. Zhu, and K. Behnia, Phys. Rev. Lett. **119**, 056601 (2017).

[24]K. Kuroda, T. Tomita, M.-T. Suzuki, C. Bareille, A. A. Nugroho, P. Goswami, M. Ochi, M. Ikhlas, M. Nakayama, S. Akebi, R. Noguchi, R. Ishii, N. Inami, K. Ono, H. Kumigashira, A. Varykhalov, T. Muro, T. Koretsune, R. Arita, S. Shin, Takeshi Kondo, and S. Nakatsuji, Nat. Mater. **16**, 1090 (2017).

[25]G. Sharma, P. Goswami, and S. Tewari, Phys. Rev. B **93**, 035116 (2016).

[26]A. Sakai, Y.P. Mizuta, A.A. Nugroho, R. Sihombing, T. Koretsune, M.-T. Suzuki, N. Takemori, R. Ishii, D. Nishio-Hamane, R. Arita, P. Goswami, and S. Nakatsuji, Nat. Phys. **14**, 1119 (2018).

[27]S. N. Guin, K. Manna, J. Noky, S. J. Watzman, C. Fu, N. Kumar, W. Schnelle, C. Shekhar, Y. Sun, J. Gooth, and C. Felser, NPG Asia Materials **11**, 16 (2019).

[28]J. Kübler, and C. Felser, Europhys. Lett. **114**, 47005 (2016).

[29]G. Chang, S.-Y. Xu, X. Zhou, S.-M. Huang, B. Singh, B. Wang, I. Belopolski, J. Yin, S. Zhang, A. Bansil, H. Lin, and M. Z. Hasan, Phys. Rev. Lett. **119**, 156401 (2017).

[30]K. Sumida, Y. Sakuraba, K. Masuda, T. Kono, M. Kakoki, K. Goto, W. Zhou, K. Miyamoto, Y. Miura, T. Okuda, and A. Kimura, Commun. Mater. **1**, 89 (2020).

[31]H. Reichlova, R. Schlitz, S. Beckert, P. Swekis, A. Markou, Y.C. Chen, D. Kriegner, S. Fabretti, G. Hyeon Park, A. Niemann, S. Sudheendra, A. Thomas, K. Nielsch, C. Felser, and S.T.B. Goennenwein, Appl. Phys. Lett. **113**, 212405 (2018).

[32]Y. Sakuraba, K. Takanashi, Y. Kota, T. Kubota, M. Oogane, A. Sakuma, and Y. Ando, Phys. Rev. B **81**, 144422 (2010).

[33]Y. Sakuraba, S. Kokado, Y. Hirayama, T. Furubayashi, H. Sukegawa, SFarea. Li, Y. K. Takahashi, and K. Hono, Appl. Phys. Lett. **104**, 172407 (2014)

[34]Y. Sakuraba, K. Hyodo, A. Sakuma, and S. Mitani, Phys. Rev. B **101**, 134407 (2020).

[35]S. Ueda, Y. Katsuya, M. Tanaka, H. Yoshikawa, Y. Yamashita, S. Ishimaru, Y. Matushita and K. Kobayashi, AIP Conf. Proc. **1234**, 403 (2010).

[36]S. Ueda, J. Electron Spectrosc. Rel. Phenom. **190**, 235 (2013).

[37]G. Kresse and J. Hafner, Phys. Rev. B **47**, 558 (1993).

[38]G. Kresse G and J. Furthmüller, Comput. Mater. Sci. **6**, 15 (1996).

[39]J. P. Perdew, K. Burke and M. Ernzerhof, Phys. Rev. Lett. **77**, 3865 (1996).

[40]P. E. Blöchl, Phys. Rev. B **50**, 17953 (1994).

[41]G. Kresse and D. Joubert, Phys. Rev. B **59**, 1758 (1999).

[42]K. Uchida, S. Daimon, R. Iguchi, and E. Saitoh, Nature **558**, 95 (2018).

[43]S. Daimon, R. Iguchi, T. Hioki, E. Saitoh, and K. Uchida, Nat. Commun. **7**, 13754 (2016).

[44]S. Daimon, K. Uchida, R. Iguchi, T. Hioki, and E. Saitoh, Phys. Rev. B **96**, 024424 (2017).

[45]O. Wid, J. Bauer, A. Muller, O. Breitenstein, S. S. P. Parkin, and G. Schmidt, Sci. Rep. **6**, 28233 (2016).

[46]O. Breitenstein, W. Warta, and M. Langenkamp, Lock-in thermography: Basics and Use for Evaluating Electronic Devices and Materials Introduction. (Springer, Berlin/Heidelberg, Germany, 2010).

[47]T. Seki, R. Iguchi, K. Takanashi, and K. Uchida, Appl. Phys. Lett. **112**, 152403 (2018).

[48]T. Seki, R. Iguchi, K. Takanashi, and K. Uchida, J. Phys. D **51**, 254001 (2018).

[49]T. Seki, A. Miura, K. Uchida, T. Kubota, and K. Takanashi, Appl. Phys. Express **12**, 023006 (2019).

[50]S. Ota, K. Uchida, R. Iguchi, P. Van Thach, H. Awano, and D. Chiba, Sci. Rep. **9**, 13197 (2019).

[51] A. Miura, H. Sepehri-Amin, K. Masuda, H. Tsuchiura, Y. Miura, R. Iguchi, Y. Sakuraba, J. Shiomi, K. Hono, and K. Uchida, Appl. Phys. Lett. **115**, 222403 (2019).

[52] R Modak, and K. Uchida, Appl. Phys. Lett. **116**, 032403 (2020).

[53] A. Miura, R. Iguchi, T. Seki, K. Takanashi, and K. Uchida, Phys. Rev. Materials **4**, 034409 (2020).

[54] Y. Takamura, R. Nakane, and S. Sugahara, J. Appl. Phys. **105**, 07B109 (2009).

[55] R. Y. Umetsu, K. Kobayashi, A. Fujita, R. Kainuma, and K. Ishida, Scr. Mater. **58**, 723 (2008).

[56] S. Ouardi, A. Gloskovskii, B. Balke, C. A. Jenkins, J. Barth, G. H. Fecher, C. Felser, M. Gorgoi, M. Mertin, F. Schafers, E. Ikenaga, K. Yang, K. Kobayashi, T. Kubota, M. Oogane, and Y, Ando, J. Phys. D: Appl. Phys. **42**, 084011 (2009).

[57] B. Balke, G. H. Fecher, H. C. Kandpal, C. Felser, K. Kobayashi, E. Ikenaga, J.-J. Kim, and S. Ueda, Phys. Rev. B **74**, 104405 (2006).

[58] G. H. Fecher, B. Balke, A. Gloskowskii, S. Ouardi, C. Felser, T. Ishikawa, M. Yamamoto, Y. Yamashita, H. Yoshikawa, S. Ueda, and K. Kobayashi, Appl. Phys. Lett. **92**, 193513 (2008).

[59] G. H. Fecher, B. Balke, S. Ouardi, C. Felser, G. Schonhense G, E. Ikenaga, J.-J. Kim, S. Ueda and K. Kobayashi K, J. Phys. D: Appl. Phys. **40**, 1576 (2007).

[60] P. Li, J. Koo, W. Ning, J. Li, L. Miao, L. Min, Y. Zhu, Y. Wang, N. Alem, C.-X. Liu, Z. Mao, and B. Yan, Nat. Commun. **11**, 3476 (2020).

[61] K. Uchida, M. Sasaki, Y. Sakuraba, R. Iguchi, S. Daimon, E. Saitoh, and M. Goto, Sci. Rep. **8**, 16067 (2018).

[62] H. Masuda, R. Modak, T. Seki, K. Uchida, Y.-C. Lau, Y. Sakuraba, R. Iguchi, and K. Takanashi, Commun. Mater. **1**, 75 (2020).

**FIGURES**

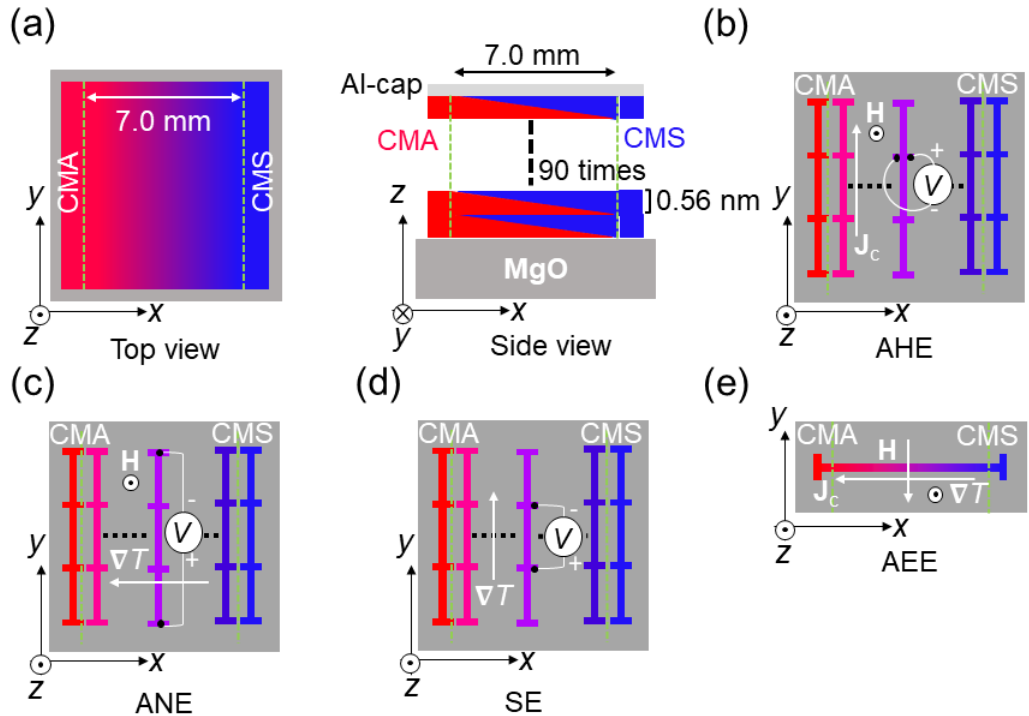

**FIG. 1.** (a) Schematic representation of 7.0 mm composition-spread combinatorial film deposited with wedge shaped layer by layer deposition. An identical film with 10 mm composition spread was deposited for HAXPES measurement. To measure the composition dependence transport and thermoelectric properties the 7.0 mm composition spread film was patterned in Hall bars perpendicular to the composition gradient. Schematic measurement configuration used for (b) AHE (c) ANE (d) SE measurement. (e) AEE measurement configuration. To measure composition dependence AEE the film is patterned in 8.0 mm long and 0.4 mm width strip along composition gradient.

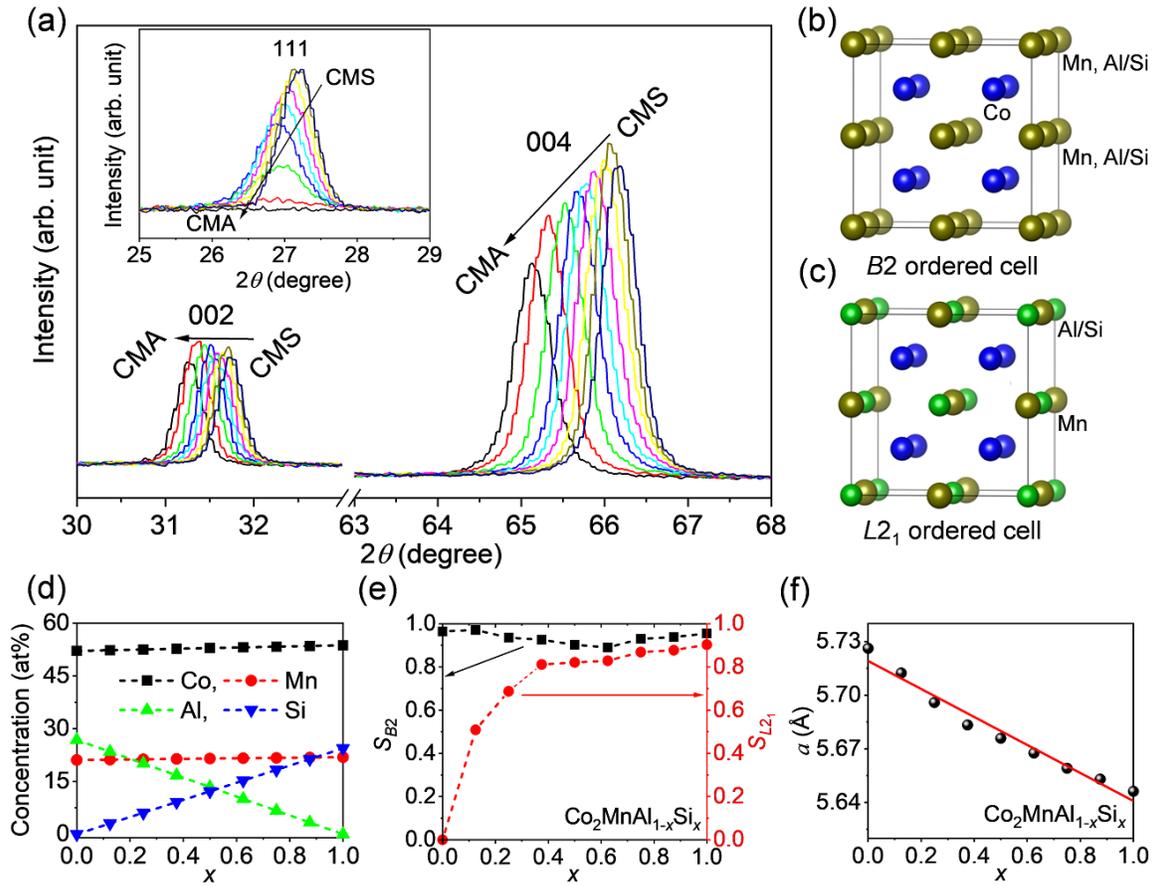

**FIG. 2.** (a) Out-of-plane X-ray diffraction patterns with x variation for $Co_2MnAl_{1-x}Si_x$ thin film of composition gradient over 7.0 mm. The measurement is performed at 1.0 mm intervals along the composition gradient using 0.5 mm incident slit (b) Schematic representation of B2 ordered unit cell for $Co_2Mn(Al/Si)$ and (c) Schematic representation of L2₁ ordered unit cell for $Co_2Mn(Al/Si)$ Heusler alloy. (d) *x* variation of estimated composition for $Co_2MnAl_{1-x}Si_x$ film with 7.0 mm composition variation. The variation is estimated from the measured composition in $x = 0$ and 1 regions in the CMAS composition-spread film and considering linear variation over the composition gradient. (e) *x* variation of $B2$ and $L2_1$ ordering coefficients for the $Co_2MnAl_{1-x}Si_x$ film with 7.0 mm composition variation. (f) *x* dependence of lattice constant (*a*) value calculated from 004 peak."

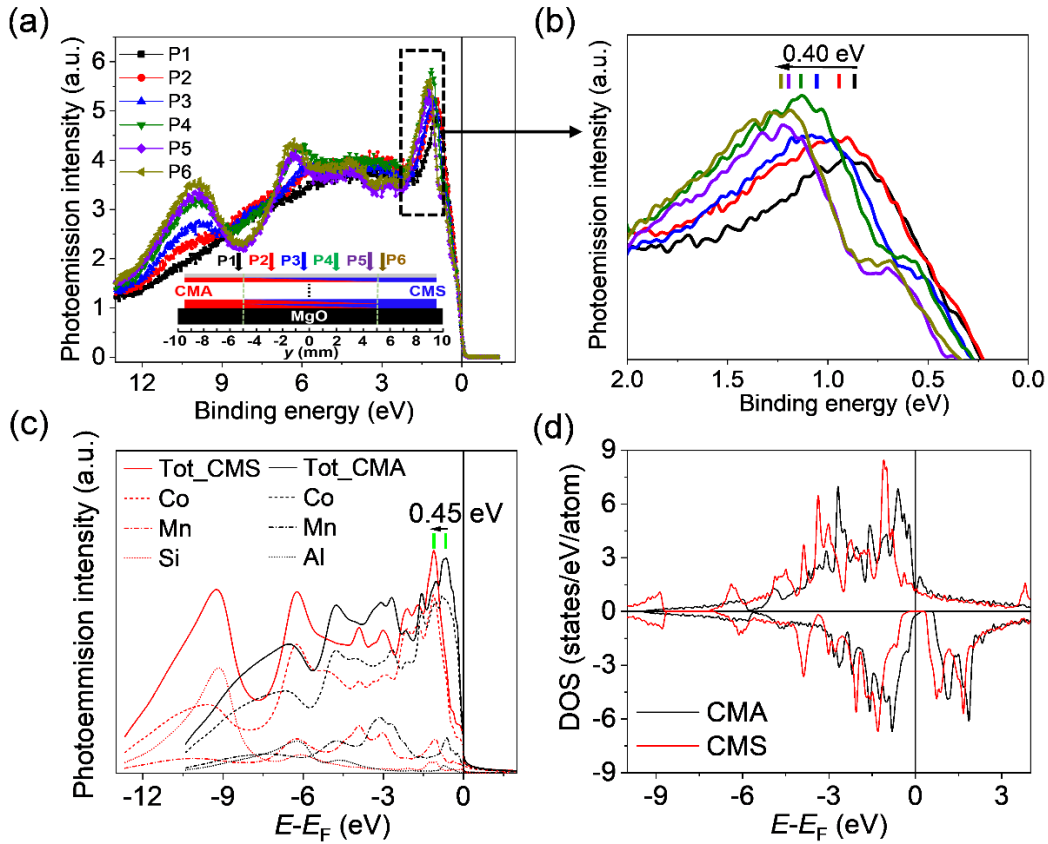

**FIG. 3.** (a) Valance band HAXPES spectra for CMAS composition-spread film recorded at different positions (P1 to P6). (b) The enlarged view of the spectra near Fermi level (c) Simulated HAXPES spectra for $Co_2MnSi$ and $Co_2MnAl$ alloy considering perfect $L2_1$ ordering unit cell (d) Calculated spin-resolved DOSs for $L2_1$-ordered $Co_2MnSi$ and $Co_2MnAl$ using experimental lattice constant.

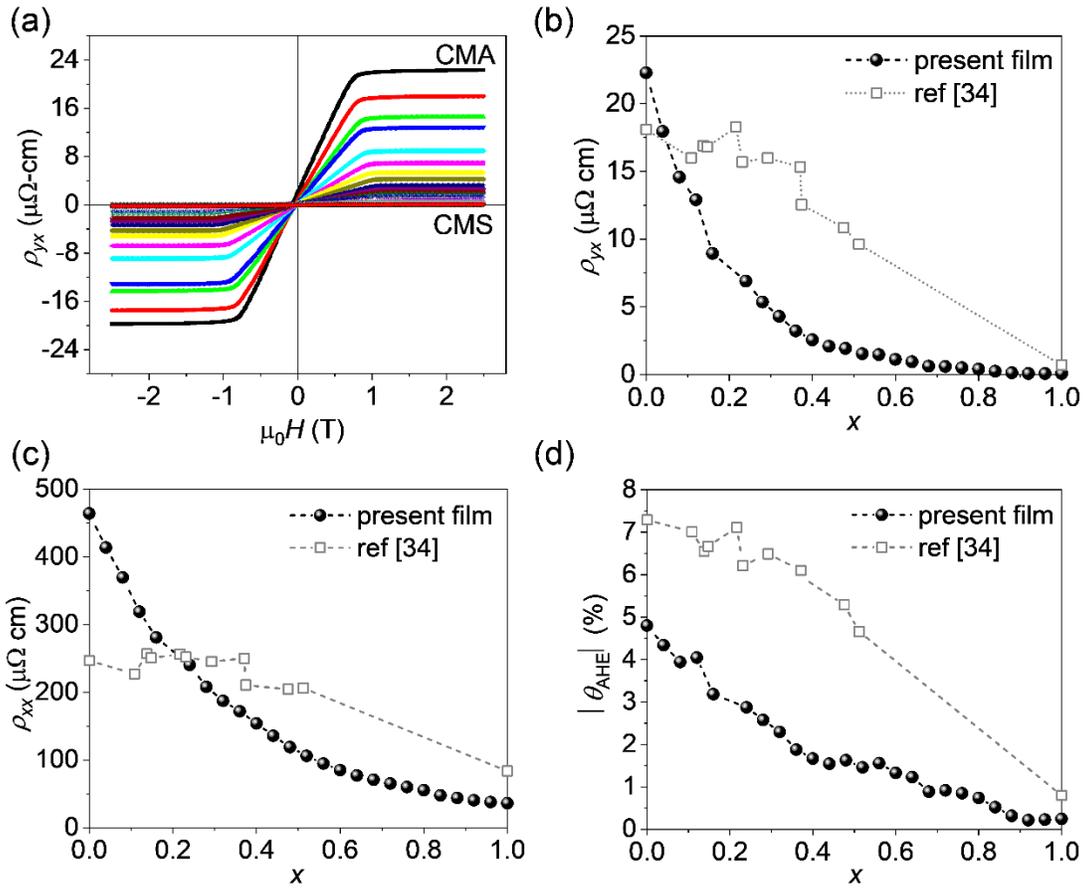

**FIG. 4.** (a) Perpendicular magnetic field $H$ dependence of anomalous Hall resistivity ($\rho_{yx}$) for $Co_2MnAl_{1-x}Si_x$ combi film (with 7.0 mm composition gradient) for different $x$ values. $x$ (Si percentage) dependence of (b) $\rho_{yx}$ (c) electrical resistivity ($\rho_{xx}$) and (d) anomalous Hall angle ($\theta_{AHE}$) for $Co_2MnAl_{1-x}Si_x$ combi film along the composition gradient.

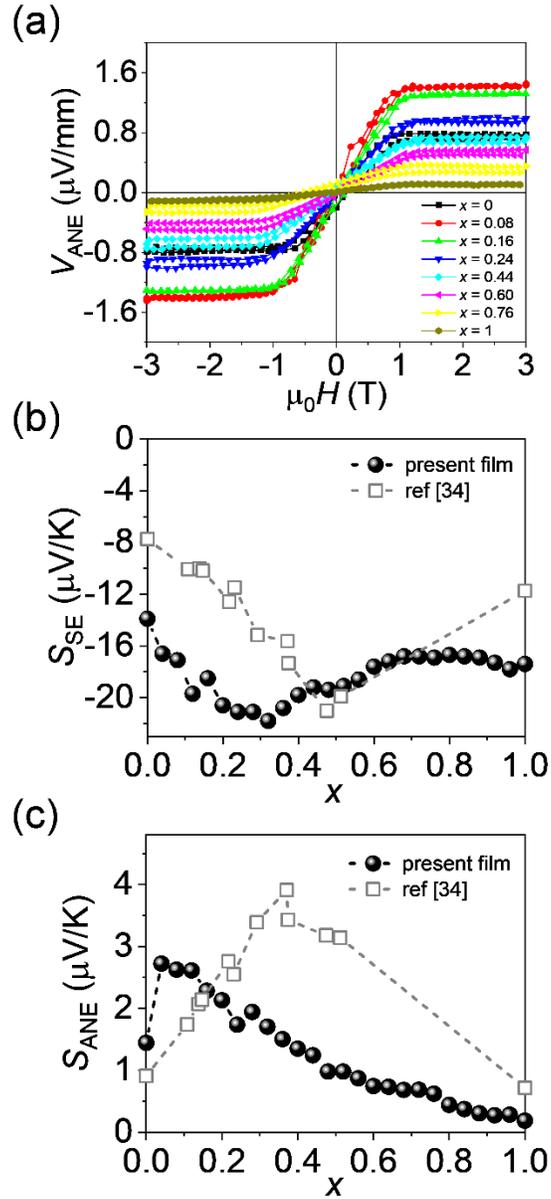

**FIG. 5.** (a) Perpendicular magnetic field $H$ dependence of anomalous Nernst voltage $V_{ANE}$ for $Co_2MnAl_{1-x}Si_x$ combi film (with 7.0 mm composition gradient) for different $x$ values (b) $x$ (Si composition) dependence of (b) Seebeck coefficient ($S_{SE}$) and (c) anomalous Nernst coefficient ($S_{ANE}$).

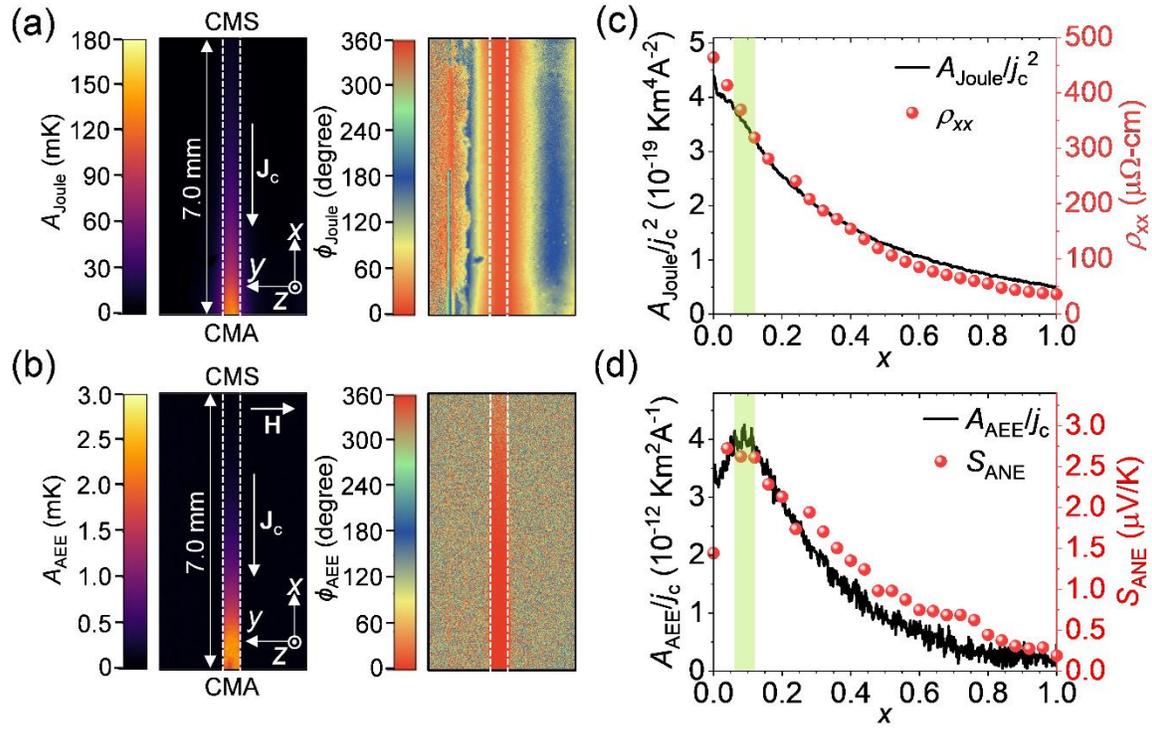

**FIG. 6.** (a) Amplitude ($A_{\text{Joule}}$) and phase ($\phi_{\text{Joule}}$) images of the Joule-heating-induced temperature modulation along the composition gradient (7.0 mm long) for the frequency $f = 25$ Hz and applied charge current $J_c = 10$ mA, which corresponds to the charge current density $j_c = 4.96 \times 10^8$ Am$^{-2}$. (b) Amplitude ($A_{\text{AEE}}$) and phase ($\phi_{\text{AEE}}$) images of AEE-induced temperature modulation for $f = 25$ Hz, $J_c = 10$ mA, and $\mu_0 H = 0.1$ T (along film plane). (c) $A_{\text{Joule}}/j_c^2$ and $\rho_{xx}$ as a function of $x$. (d) $A_{\text{AEE}}/j_c$ and $S_{\text{ANE}}$ as a function of $x$.

# Supplementary Material

## I. Procedure for extracting pure AEE signal:

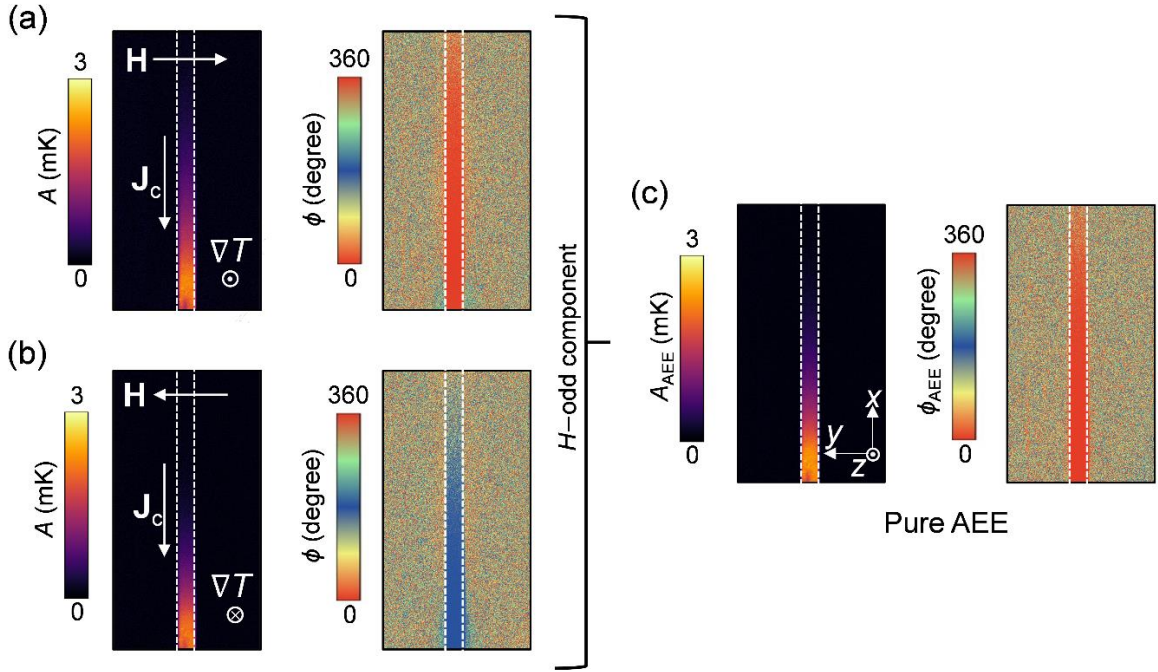

**Fig. S1.** (a,b) Raw lock-in amplitude $A$ and phase $\phi$ images for two different field directions. $\mathbf{J}_c$ is the applied charge current, $\mathbf{H}$ is the external magnetic field, and $\nabla T$ is the temperature gradient induced by AEE. (c) $A_{\text{AEE}}$ and $\phi_{\text{AEE}}$ images showing the pure AEE signal, obtained by subtracting the images in (b) from those in (a).

The temperature modulation induced by AEE for the composition-spread film was measured by using the lock-in thermography (LIT) technique.[1-5] Figs. S1(a) and (b) show the raw lock-in amplitude ($A$) and phase ($\phi$) images of the temperature modulation generated from the square-wave-modulated AC charge current with zero DC offset, the amplitude of $J_c$ = 10 mA, and the frequency of $f$ = 25 Hz, measured when the charge current was applied along the $x$-direction and the magnetic field of 0.1 T was applied along the $-y$ or $+y$-directions. Here, $A$ represents the magnitude of the current-induced temperature modulation, while $\phi$ represents the sign of the temperature modulation at the top surface of the sample when the time delay of the temperature modulation due to thermal diffusion is negligible.[6] In this configuration, the temperature gradient $\nabla T$ is generated along the $z$-direction following the symmetry of AEE:[6-12]

$$\nabla T \propto \mathbf{J}_c \times \mathbf{M}$$

where $\mathbf{J}_c$ is the charge current and $\mathbf{M}$ is the magnetization direction in a ferromagnet. The ~180° phase change between the images in Fig. S1(a) and (b) clearly demonstrates the symmetry of AEE because $\mathbf{M}$ is reversed by reversing the magnetic field, indicating the presence of the dominant AEE contribution in the obtained thermal images. However, the raw images may also contain small field-independent thermoelectric signals due to the Peltier effect. To extract the pure AEE signals, the lock-in amplitude $A_{AEE}$ and phase $\phi_{AEE}$ images showing the field-odd dependence were obtained by subtracting the raw LIT images at $\mu_0 H = -0.1$ T [Fig. S1(b)] from those at $\mu_0 H = +0.1$ T [Fig. S1(a)] and dividing the subtracted images by 2.[6-12] Figure S1(c) represents the subtracted images that are used for the analysis in the main text.

**II. Procedure of XRD pattern simulation:**

| x | Elementals (at%) | | | | Simulated XRD Intensity | | |
|---|---|---|---|---|---|---|---|
| | Co | Mn | Al | Si | $I_{111}$ | $I_{002}$ | $I_{004}$ |
| 0.000 | 52.10 | 21.10 | 26.80 | 0.00 | 5.53 | 4.10 | 12.82 |
| 0.125 | 52.30 | 21.20 | 23.45 | 3.05 | 5.55 | 3.94 | 12.80 |
| 0.250 | 52.50 | 21.30 | 20.10 | 6.10 | 5.58 | 3.78 | 12.78 |
| 0.375 | 52.70 | 21.40 | 16.75 | 9.15 | 5.60 | 3.63 | 12.76 |
| 0.500 | 52.90 | 21.50 | 13.40 | 12.20 | 5.61 | 3.48 | 12.75 |
| 0.625 | 53.10 | 21.60 | 10.05 | 15.25 | 5.63 | 3.33 | 12.74 |
| 0.750 | 53.30 | 21.70 | 6.70 | 18.30 | 5.64 | 3.19 | 12.72 |
| 0.875 | 53.50 | 21.80 | 3.35 | 21.35 | 5.52 | 3.14 | 12.72 |
| 1.000 | 53.70 | 21.90 | 0.00 | 24.40 | 5.41 | 3.09 | 12.71 |

**Table S1**: Shows the estimated composition values used to simulate XRD pattern using visualization for electronic and structural analysis (VESTA) and corresponding normalized intensities for 111, 002 and 004 peaks. To generate XRD we have constructed a perfect $L2_1$ ordered Heusler unit cell of chemical formula $X_2YZ$ with X, Y and Z occupied 8c, 4b and 4a Wykoff positions. For the present case we have used the actual film composition with modified structure as $X_{2+a}Y_{1-a-b}Z_{1+b}$ (X = Co, Y = Mn, Z = Si/Al), where *a*, *b* represents the additional occupancy due to the deviation of actual composition from stoichiometry. For simplicity, we have considered the following: X position is completely occupied by Co, Z position is completely occupied by Si/Al and the Y position is occupied by Mn and excess Co and Al/Si.

## III. Transport and thermoelectric properties of Co$_2$MnAl$_{1-x}$Si$_x$ film:

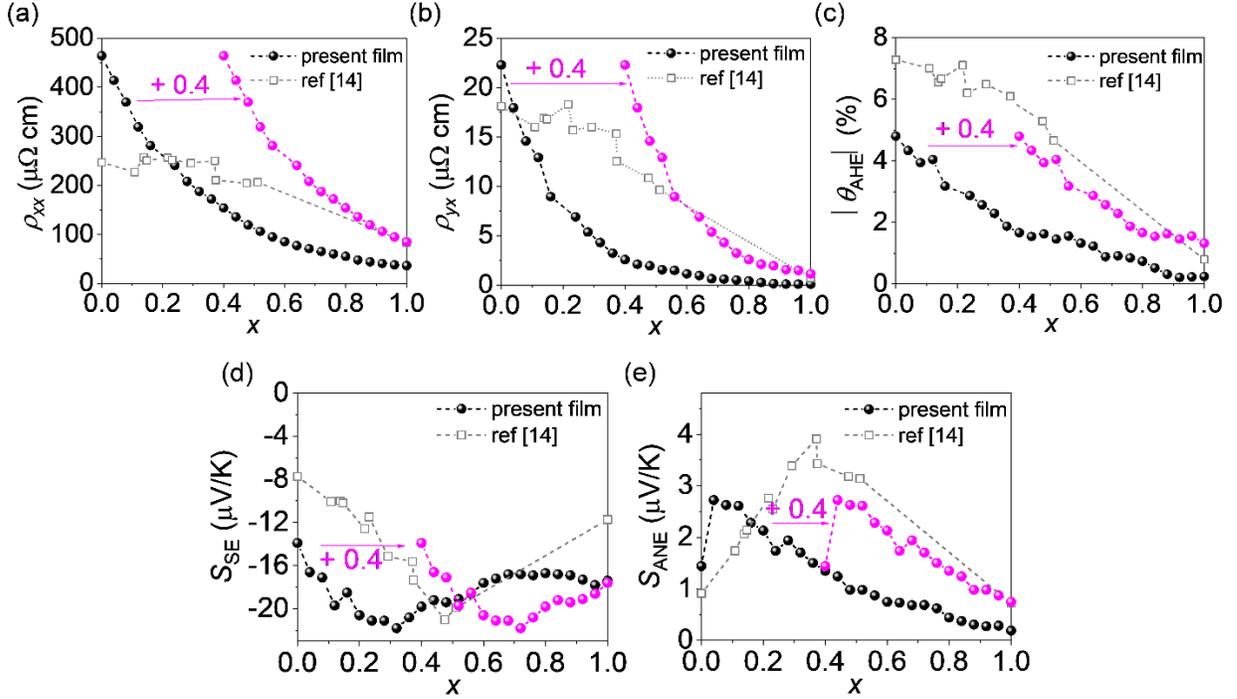

**Fig. S2.** $x$ (Si atomic ratio to Al) dependence of (a) electrical resistivity ($\rho_{xx}$), (b) anomalous Hall resistivity ($\rho_{yx}$) (c) anomalous Hall angle ($\theta_{AHE}$), (d) Seebeck coefficient ($S_{SE}$), and (e) anomalous Nernst coefficient ($S_{ANE}$) for the Co$_2$MnAl$_{1-x}$Si$_x$ (CMAS) composition-spread film along the composition gradient. The data shown in pink circle is the data with +0.4 offset to the present CMAS film to adjust the total $N_V$ based on the $N_V$ for the Co$_2$MnAl, i.e.., $N_V$ = 27.9 and 27.5 for the present and previous films in ref. 14, respectively (ref. 14 is same as ref. 34 in the main manuscript).

Figs. S2(a) – 3(e) represent the Si atomic ratio dependence transport and thermoelectric properties for the present Co$_2$MnAl$_{1-x}$Si$_x$ (CMAS) composition-spread film (black filled symbol) compared with the Co$_2$MnAl$_{1-x}$Si$_x$ films used in ref. 14 (gray open symbol) deposited at different compositions one by one. A small difference between the present and previous films can be observed for all these properties which can be understood with the fact that the compositions for these two films are little different. For instance, the compositions for present CMA film and in ref. 14 are $Co_{52.1}Mn_{21.1}Al_{26.8}$ and $Co_{48.4}Mn_{24.5}Al_{27.1}$, respectively which corresponds to the total valance electron number $N_V$ of 27.9 and 27.5, respectively. As previous study reported on $Co_2MnGa$,[13] various electric and thermoelectric properties well scale with $N_V$. Similar behavior is also expected for CMAS films, which is reflected in the present study. The $N_V$ value for CMA (at $x = 0$ for present film) is higher by 0.4 as compared to CMA in ref. 14 due to the change in composition. As can be seen in the above figure (magenta filled symbol) if we consider a shift of the present data to an equivalent $N_V$ value of the previously reported films in ref. 14 (around +0.4), the electric and thermoelectric properties including $S_{SE}$ and $S_{ANE}$ show a closer matching although there are still deviations that are expected to arise from a several factors that do not scale with only $N_V$ such as the degree of $L2_1$ atomic order. Here it should be noted that till now there is no clear report on off-stoichiometry composition or $N_V$ dependence of transport properties for CMAS films. The present result suggests a strong correlation between them.

## IV. Band structure calculation for Co₂MnAl and Co₂MnSi alloy:

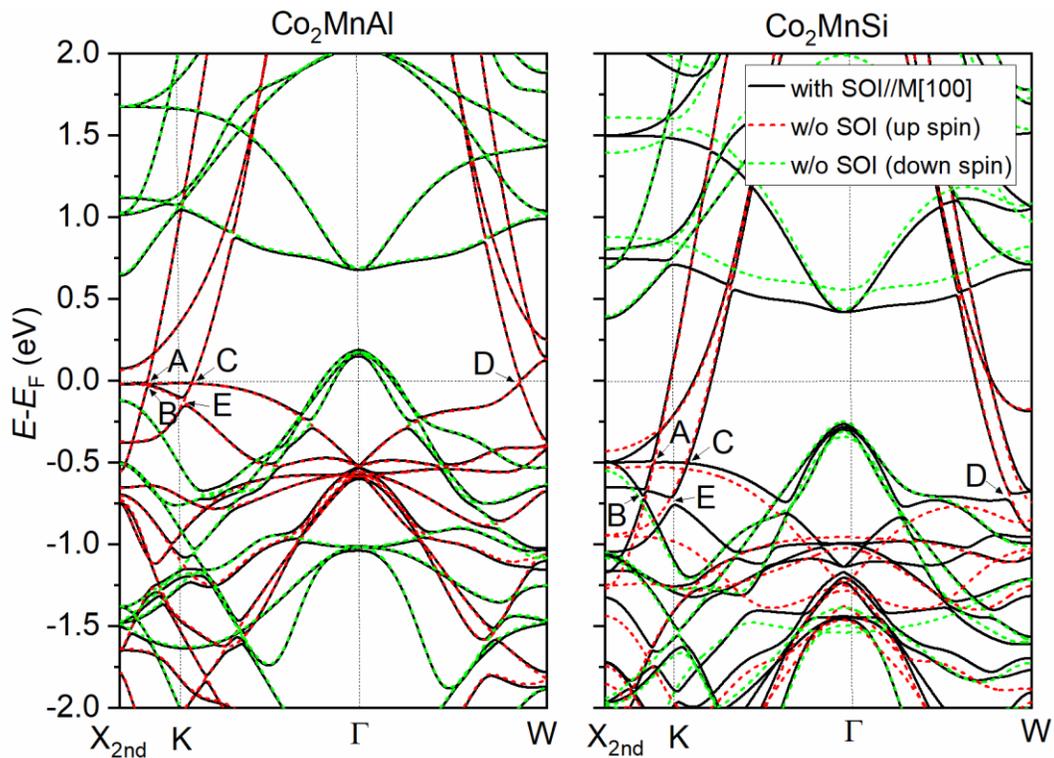

**Fig. S3.** Calculated band structures for $L2_1$ ordered Co₂MnAl and Co₂MnSi alloys. The Weyl cones are indexed as A to E by following ref.13. The calculation clearly demonstrates the shifting of energy level with Si substitution.

**References:**

[1] K. Uchida, S. Daimon, R. Iguchi, and E. Saitoh, Nature **558**, 95 (2018).

[2] S. Daimon, R. Iguchi, T. Hioki, E. Saitoh, and K. Uchida, Nat. Commun. **7**, 13754 (2016).

[3] S. Daimon, K. Uchida, R. Iguchi, T. Hioki, and E. Saitoh, Phys. Rev. B **96**, 024424 (2017).

[4] O. Wid, J. Bauer, A. Muller, O. Breitenstein, S. S. P. Parkin, and G. Schmidt, Sci. Rep. **6**, 28233 (2016).

[5] O. Breitenstein, W. Warta, and M. Langenkamp, Lock-in thermography: Basics and Use for Evaluating Electronic Devices and Materials Introduction. (Springer, Berlin/Heidelberg, Germany, 2010).

[6] T. Seki, R. Iguchi, K. Takanashi, and K. Uchida, Appl. Phys. Lett. **112**, 152403 (2018).

[7] T. Seki, R. Iguchi, K. Takanashi, and K. Uchida, J. Phys. D **51**, 254001 (2018).

[8] T. Seki, A. Miura, K. Uchida, T. Kubota, and K. Takanashi, Appl. Phys. Express **12**, 023006 (2019).

[9] S. Ota, K. Uchida, R. Iguchi, P. Van Thach, H. Awano, and D. Chiba, Sci. Rep. **9**, 13197 (2019).

[10] A. Miura, H. Sepehri-Amin, K. Masuda, H. Tsuchiura, Y. Miura, R. Iguchi, Y. Sakuraba, J. Shiomi, K. Hono, and K. Uchida, Appl. Phys. Lett. **115**, 222403 (2019).

[11] R Modak, and K. Uchida, Appl. Phys. Lett. **116**, 032403 (2020).

[12] A. Miura, R. Iguchi, T. Seki, K. Takanashi, and K. Uchida, Phys. Rev. Materials **4**, 034409 (2020).

[13] K. Sumida, Y. Sakuraba, K. Masuda, T. Kono, M. Kakoki, K. Goto, W. Zhou, K. Miyamoto, Y. Miura, T. Okuda, and A. Kimura, Commun. Mater. **1**, 89 (2020).

[14] Y. Sakuraba, K. Hyodo, A. Sakuma, and S. Mitani, Phys. Rev. B **101**, 134407 (2020).